# Ultra-narrow Linewidth Fiber Laser with Self-injection Feedback Based on Rayleigh Backscattering

Tao Zhu, *Member, IEEE*, Shihong Huang, Leilei Shi, Wei Huang, Lei Gao, Baomei Zhang

***Abstract*—A single longitudinal mode fiber laser with ultra-narrow linewidth based on self-injection feedback by using the linewidth compress mechanism of Rayleigh backscattering (RBS) are proposed and demonstrated. Since the linewidth of RBS is narrower than that of the incident light in optical fibers and they have the same centre wavelength, the RBS can act as a mechanism to compress the linewidth of the incident light in fiber ring laser. In addition, more RBS signal could be collected to help further compress the laser linewidth besides the free spectral range is expanded when the self-injection feedback method is used. Our experimental results show that the side-mode suppression ratio of our laser is up to 75dB and the laser linewidth could be low to ~130Hz.**

## I. Introduction

Single longitudinal mode (SLM) fiber laser with ultra-narrow linewidth has advantages of ultra-long coherent length and low phase noise, which can be widely used in optical fiber communications, optical fiber sensing with ultra-precision and ultra-long-range detection, optical clock. It can also be used as a seed laser to create the excellent microwave signal or as an optical frequency standard to measure the unknown light frequency accurately [1-4]. Up to now, several techniques are reported to realize SLM fiber laser, such as saturable absorber [5-7], short cavity [1, 4, 8-9], mode selection elements [10-11], passive multiple ring cavities [12], or self-injection feedback [13] and so on. The 3dB linewidth of the laser can be compressed to ~1-2 kHz by using the above mentioned methods. So it is still a challenge to compress the laser linewidth to less than 1 kHz.

In our paper, we propose a kind of Rayleigh backscattering (RBS) based mechanism to realize the modes selection and linewidth compressing. The principle relies on the fact that the linewidth of RBS is narrower than that of the incident light in optical fibers by optimizing the input power [14]. Hence, the output laser linewidth could be compressed to several hundreds of Hertz after the RBS circulates the ring with amplification hundreds of thousands times. To further compress the linewidth, we use self-injection feedback to collect more RBS, which also expands the free spectral range (FSR). Finally, the fiber laser with ~130Hz linewidth and stable output power is realized experimentally.

This work was supported by Natural Science Foundation of China (No. 61377066), the Fundamental Research Funds for the Central Universities (No. CDJZR12125502 and 106112013CDJZR120002).

Tao Zhu, Shihong Huang, Wei Huang, Baomei Zhang are with the Key Laboratory of Optoelectronic Technology & Systems (Education Ministry of China), Chongqing University, Chongqing 400044, China. (Corresponding email: zhutao@cqu.edu.cn)

## II. Principle of Laser Linewidth Compression

Rayleigh scattering occurs when the wavelength of the incident light is longer than the radius of particles. And the intensity of Rayleigh scattering light is proportional to the intensity but inversely proportional to the fourth power of wavelength of the incident light. The profile of intensity versus wavelength of incident light, corresponding RBS before and after amplification is shown in Fig.1, where $I(\lambda_0)_i$ is the incident light with Gauss or Lorentz line-shape, $I(\lambda)_s$ is the RBS, and $I(\lambda)_a$ is the amplified RBS light.

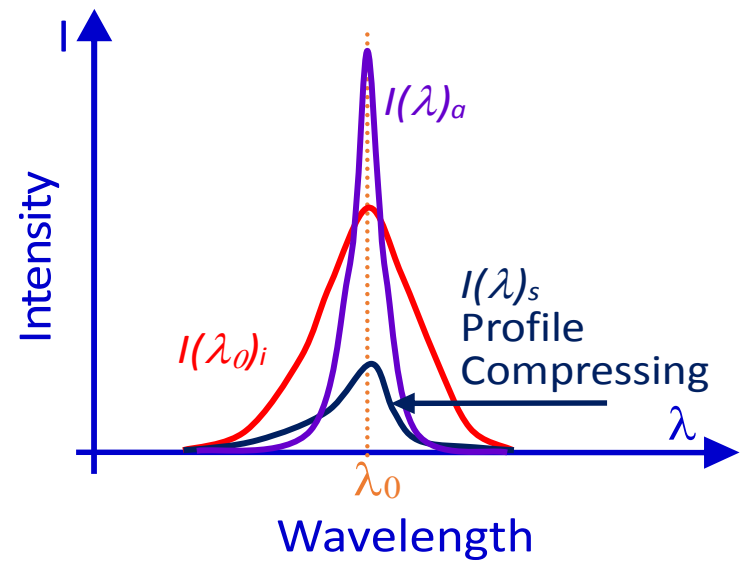

Fig.1. Schematic diagram of the relationship among the intensity of incident light, the intensities of the corresponding RBS before and after amplification.

Compared with the bulk medium, optical fiber has an enhancement nonlinear effects factor of $10^9$ due to its low attenuation coefficient and small light field diameter [15]. Therefore, RBS light could be easily accumulated and results in stimulated Rayleigh scattering [16]. When the incident light $I(\lambda_0)_i$ is injected into optical fiber with a length of $L$, the profile of RBS light $I(\lambda)_s$ located at the region $\lambda>\lambda_0$ decreases faster than that of $\lambda<\lambda_0$ since the intensity of RBS is inversely proportional to the fourth power of wavelength of the incident light, compressing the linewidth of the incident light [15]. Then RBS light $I(\lambda)_s$ will be amplified when it passes through one section of the active fiber (say Erbium doped fiber, EDF) in the fiber ring laser. And the amplified light $I(\lambda)_a$ creates RBS again in EDF with higher Rayleigh coefficient, causing narrower linewidth RBS light than that of caused by $I(\lambda)_i$. So the linewidth of output fiber ring laser will become narrower and narrower after the RBS light is excited and amplified for thousands of times. Stimulated Rayleigh excited by higher input light power will further help to compress the laser linewidth and amplify the laser power [14].

Recently, we use tapered optical fiber acting as the RBS medium to suppress stimulated Brillouin scattering and get the ultra-linewidth fiber ring laser [16]. By this way, the linewidth is compressed to 300Hz or 800Hz depending on homodyne or heterodyne measurement methods. The problems of the low pump efficiency (~0.1-0.3%) and the unstable output power still exists because of large cavity loss. In this paper, we use the self-injection feedback method to collect more RBS signal light and further compress the laser linewidth. In addition, the FSR of the fiber ring laser is expanded and the longitudinal modes can be more effectively selected, so the fiber laser with ultra-narrow linewidth, stable output can be obtained by small cavity loss. A fiber ring laser with linewidth of 100 Hz and the SMSR of 75dB achieved in our experiment. It is worth to note that such a method could also be used to compress the linewidth of other fiber lasers with different wavelength bands since Rayleigh scattering characteristics are almost the same for any of the wavelength bands.

## III. Experiments and Discussion

In general, there are two methods for measuring the linewidth of lasers, i.e., the self-heterodyne [17] and the self-homodyne [18]. Measured results about the laser linewidth is generally more accuracy by using self-homodyne method than the self-heterodyne. For the two methods, if the coherent length of the laser is longer than that of the delay fiber, the interference envelope occurs in the detection power spectrum, but if the interference envelope does not occur, we can consider that the real linewidth is narrower than that of measured.

Fig.2 (a) shows the diagram of the self-heterodyne method used in our experiments. VOA1 is used to ensure that the measured power does not change much if the pump power in the fiber laser is tuned. An acoustic optic modulator (AOM) with ~80MHz frequency shift and a normal single mode fiber are in the two arms of Mach-Zehnder interferometer, respectively. The beating signal of the two arms is detected by a photo-detector (PD) with frequency response range of DC-350MHz. Finally, the laser linewidth can be measured by the electric spectrum analyzer (ESA). PC1 and PC2 are used to make the detection result more stable. Obviously, we can get the self-homodyne measurement system when the AOM and its driver are removed from Fig.2 (a). Both of the two linewidth measurement methods will be used in our following experiments.

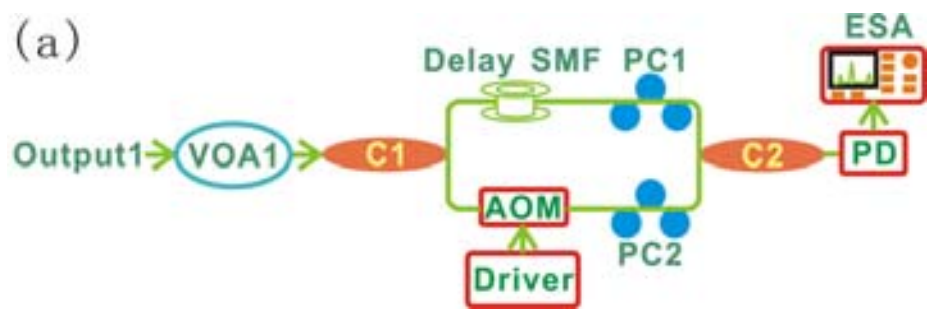

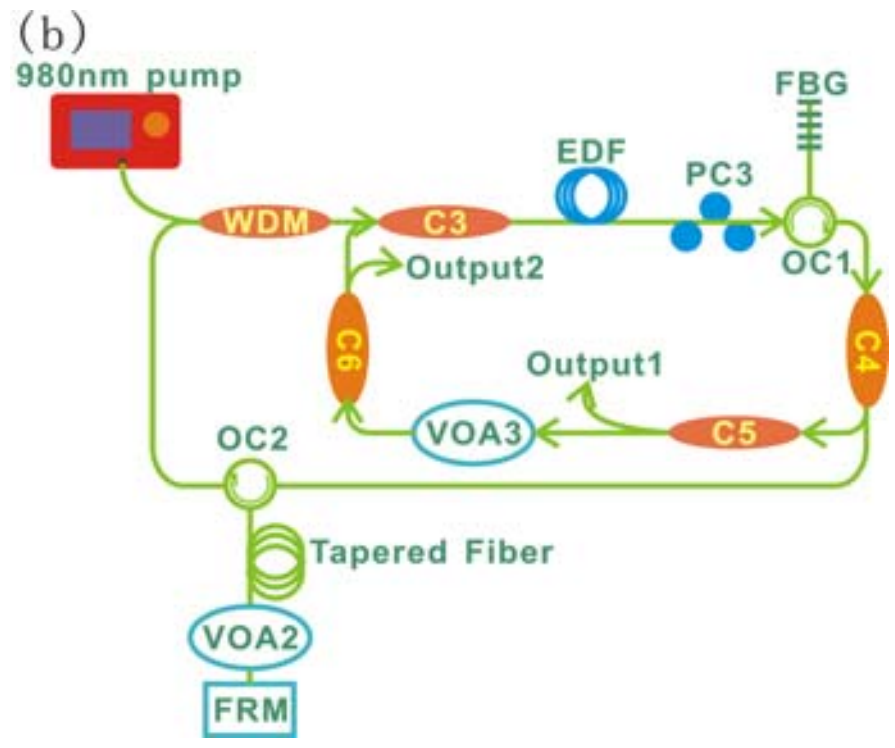

Fig. 2. (a) The self-heterodyne detection system. (b) The schematic diagram of the fiber ring laser with self-feedback based on RBS. WDM: wavelength division multiplexing device; C1: 50/50 coupler; EDF: Erbium-doped fiber; PC1, PC2, PC3: polarization controller; FBG: fiber Bragg grating; C2:50/50 coupler; OC1, OC2: optical circulator; C3: 80/20 coupler; VOA1, VOA2: variable optical attenuator; FRM: Faraday rotating mirror; C4: 50/50 coupler; AOM: acoustic-optic modulator; C5:50/50 coupler; C6:90/10 coupler; PD: photo-detector; ESA: electric spectrum analyzer.

Our experiment setup containing the self-injection feedback structure based on RBS is shown in Fig. 2(b). In the maser ring cavity, the wavelength division multiplexer (WDM), Erbium doped fiber (EDF), PC3, fiber Bragg grating (FBG), Optical circulator (OC1 and OC2) and Faraday rotation mirror (FRM) constitute a typical multiple longitudinal modes fiber ring laser. A 12 m long EDF (EDF-980-T2, Stockeryale, Inc.) with a peak absorption of 6.56 dB/m at 1530 nm is used as gain medium, which is pumped by a 980 nm laser through the WDM. The PC is used to control the state of the polarization in fiber ring to make sure the maximum output laser. OC1 and FBG with the resonant wavelength of 1550.48nm and the 3dB bandwidth of 0.2nm are used as the wavelength selection element for the fiber ring laser. Fig.3 shows output spectrum with the multiple longitudinal modes, where the FSR is ~8 MHz corresponding to the cavity length of 25m, and the linewidth of one of the longitudinal modes is ~1~2kHz.

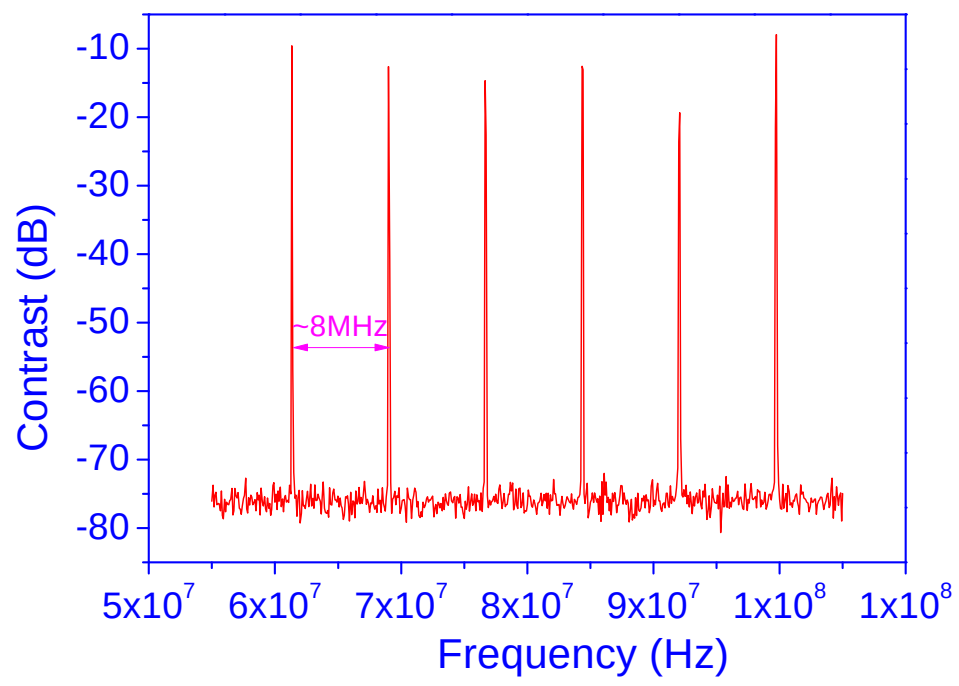

Fig.3 The output ESA spectrum of the multiple longitudinal modes fiber laser.

As shown in Fig.2 (b), when the tapered fiber with the length of 110m and VOA2 is placed between OC2 and FRM, the fiber ring laser will be injected into the slightly tapered fiber to create the narrow linewidth RBS signal through OC2. The FRM and VOA2 are used to balance the gain and the loss in the laser cavity. The slightly tapered fiber with a waist diameter of ~112μm and tapering length of ~2cm functions as the RBS structure. There are 21 tapers along the 110m single-mode fiber and the distance between two tapers is ~5m. Moreover the tapered fiber can restrict the propagation of the acoustic waves to suppress the stimulated Brillouin scattering (SBS), so more RBS signal under the larger laser power could be collected, which helps to further compress the laser linewidth.

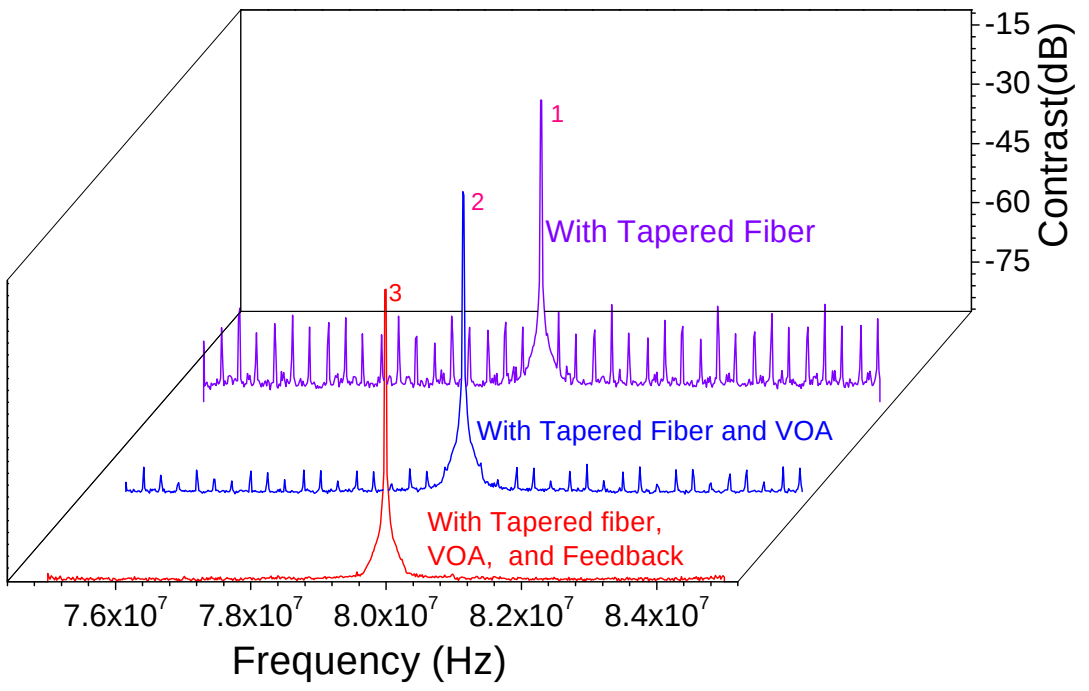

Fig.4 The longitudinal mode characteristics of the output laser without and with the self-injection feedback based on RBS.

The curve1 shown in Fig.4 indicates that modes selection function of tapered fiber caused RBS works, however, the SMSR is not good because the balance between the power from the FRM and the RBS light is not optimized. The spectrum with better SMSR (curve2, also shown in Fig.4) is achieved after we slowly adjust VOA2 under the pump power of 140mW. However, the linewidth only could be narrow to ~300Hz or ~800Hz by homodyne or heterodyne measurement methods, respectively [18]. By comparing the curve1 and the curve2, we can conclude that RBS can achieve SLM operation with good SMSR for the fiber laser, however, the better SMSR and the narrower linewidth cannot be realized except using another way.

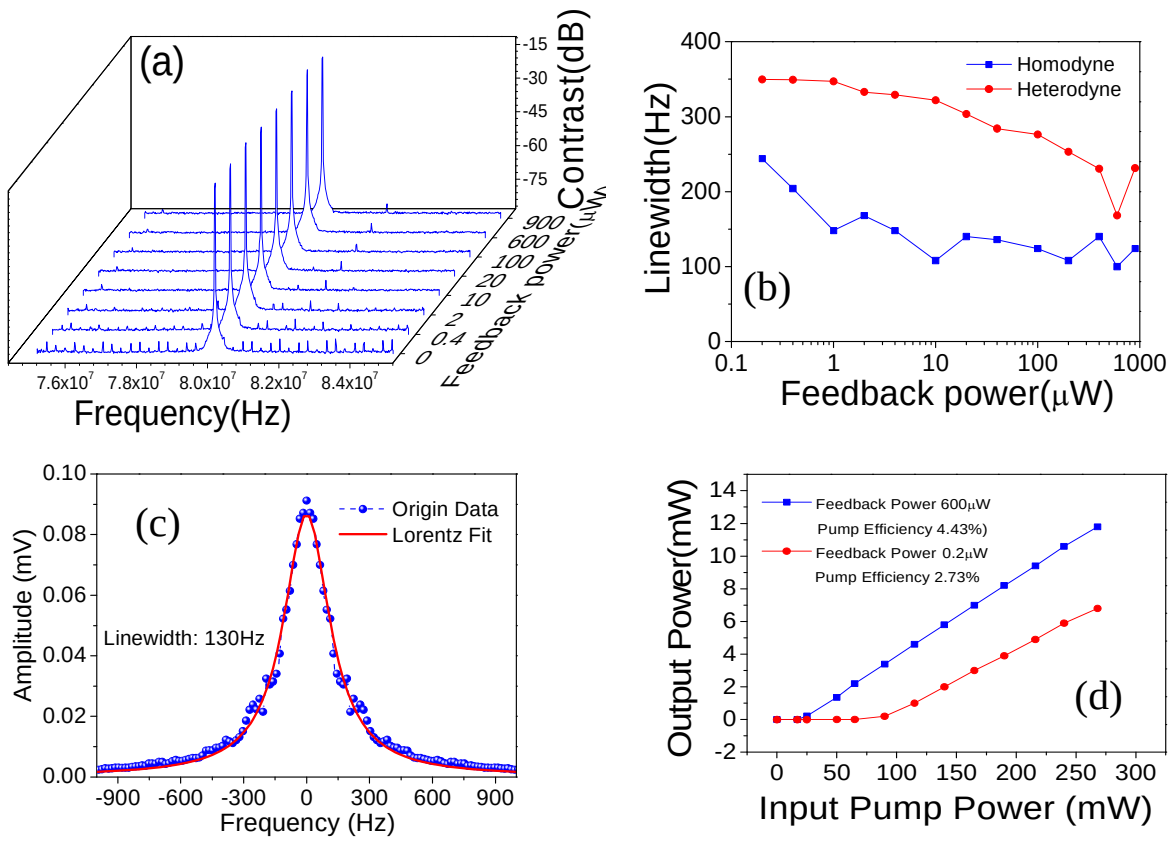

Fig.5 (a). the output spectra measured by the self-heterodyne method with the change of feedback power, (b).the relationship between the laser linewidth and the feedback power under the condition of pump power of ~165mW, (c). the Lorentz fitting linewidth for the narrowest laser linewidth, (d).the relationship between the input pump power and the output power from output1 when the feedback power is 0.2μW and 600μW, respectively.

Here we propose self-injection feedback method to compress the laser linewidth and enhance the SMSR, so C3, C5, C6, and VOA3 are added to Fig.2 (b). C4 is used to determine the laser power in the main ring cavity, and C5 is used to determine the ratio between the injection power and output laser. The feedback power could be tuned by slowly adjusting the VOA3, and is monitored by the optical power meter (OPM) through C6. The characteristic of the output laser is detected through output1. For the self-injection feedback system based on RBS, the feedback power is firstly optimized under the pump power of ~165mW. The spectrum detected by ESA is shown in Fig.5 (a) when the feedback injection power changes from 0.2μW to 900μW by adjusting VOA3 under the condition of a constant input power of ~165mW. With the increase of injection pump power, the side-mode of the output laser is suppressed gradually. Finally, many side-modes disappear when the injection power is larger than 20μW. Fig.5 (b) shows the results of Lorentz fitting linewidth under different injection power by using the delay self-homodyne and self-heterodyne measurement, respectively. We can see the output laser linewidth is compressed gradually with the increase of the feedback injection power. The output laser linewidth tends to decrease when the injection power is greater than 20μW. The narrowest linewidth of ~130Hz is achieved when the feedback power is ~600μW and the input pump power is ~165mW. The measurement result of the laser linewidth is shown in Fig. 5(c) when the feedback power is 600μW, where the linewidth is ~130Hz. And the ESA spectrum is shown in Fig.4 (Curve 3), where the SMSR is as high as ~75dB. The relationship between the input pump power and the output power from output1 is shown in Fig.5 (d) when the feedback power is 0.2μW and 600μW, respectively. We can see that the pump efficiency of the condition of 600μW is larger than that of 0.2μW. And it is much larger than 0.1~0.3% that we reported only based on RBS before [18].

The self-injection method has the function of stabling the laser frequency, so the power of the center frequency or wavelength will be more stable and could be enhanced. The power injected into the tapered fiber is also stronger than that without the feedback structure, hence the more RBS could be collected, and if possible, the stimulated Rayleigh scattering will occur. Both of the two effects help further compress the laser linewidth.

Based on the optimized feedback power of ~600μW, the optical spectra from output1 with the different input pump power is shown in Fig.6 (a). The centre wavelengths are ~1550.48nm corresponding to the resonant wavelength of FBG, but the linewidth of the laser is too narrow to be recognized by the OSA because of its resolution (~0.06nm). Fig.6 (b) shows the spectrum with the different input pump power measured by delay self-heterodyne and self-homodyne methods, respectively. The side modes of the output laser are suppressed seriously because of the broadening FSR by using the self-injection feedback, and the SMSR is up to 75dB. Then we reduce the span of ESA to measure the linewidth of the output laser with different input pump power. Fig 6(c) shows the change of the laser linewidth with the different pump power, and the fluctuation of the linewidth is in the range of 230Hz~320Hz and 100Hz ~200Hz when the pump power is in the range from 25mW to 300mW, where the delay self-heterodyne and self-homodyne measurement method are used, respectively. From the experimental results we can conclude that the linewidth compress mechanism of Rayleigh backscattering (RBS) to get the ultra-narrow linewidth laser is available by using the self-injection feedback and the tapered fiber.

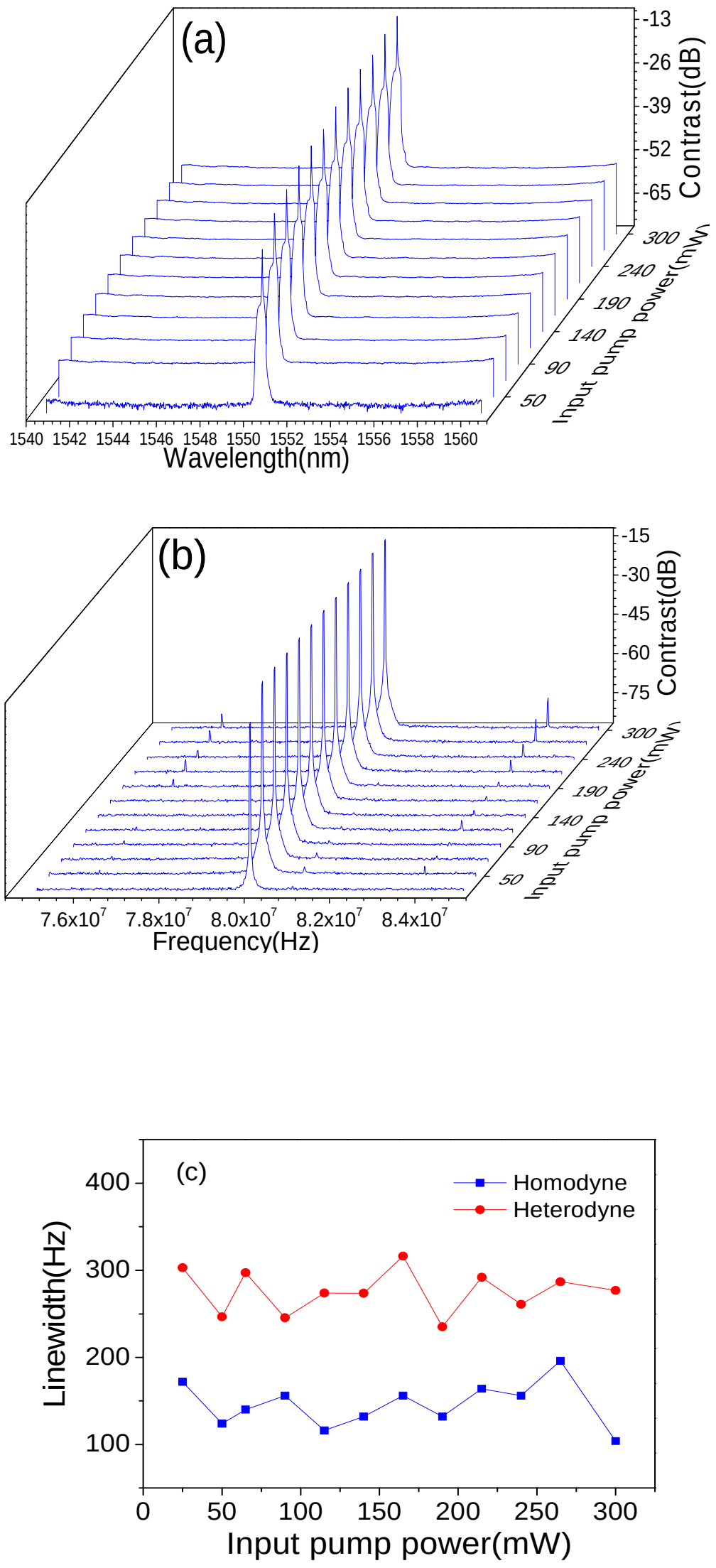

Fig.6 The experimental results when the feedback power is ~600μW. (a) the output spectrum detected by OSA, (b) the longitudinal mode characteristics detected by ESA, (c) the relationship between the laser linewidth and the different pump.

The ambient vibration should be suppressed for the ultra-narrow linewidth fiber laser. When the environment is quiet, the laser linewidth will decrease. So the antivibration structure is required for this kind of laser. However, the power fluctuation is less than 0.2dB within 48hours for our laser. Since the guided entropy could induce a little frequency shift of the RBS [19], the environmental temperature and the frequency stabilizing device should be considered. In order to make the fiber ring laser to be more compact, we need to find another fiber with larger scattering coefficient and low loss to replace the tapered fiber, such as liquid core optical fiber or pneumatic auxiliary fiber. Also we should find a method to further increase the pump efficiency in the next step.

## IV. Conclusion

In conclusion, we proposed a method to realize ultra-narrow linewidth fiber laser with stable output power and good pump efficiency using self-injection feedback and circling and amplifying RBS light. The circling and amplifying RBS light in the fiber ring laser provides a new mechanism to realize modes selection and linewidth compression because the linewidth of RBS light is lower than that of the incident light in optical fibers under optimized input power. The self-injection method ensures the high stable output power and good pump efficiency. Meanwhile, it helps to accumulate more RBS and further compress the laser linewidth. Our experimental results show that the narrowest linewidth of the output laser is ~130Hz, and the SMSR is up to 75dB. Such kind of ultra-narrow linewidth fiber laser could find potential applications in optical fiber coherent communications, optical fiber sensing with ultra-precision and ultra-long-range detection, optical clock, microwave photonics, laser radar and so on.